\documentclass[]{spie}  

\usepackage{amsmath,amsfonts,amssymb}
\usepackage{graphicx}
\usepackage[colorlinks=true, allcolors=blue]{hyperref}
\usepackage{orcidlink}
\usepackage{placeins}

\title{The Simons Observatory: Laboratory Beam Characterization for the first Small Aperture Telescope}

\author[a]{Remington G. Gerras\orcidlink{0009-0009-0876-9168}}
\author[b]{Thomas Alford\orcidlink{0000-0003-1942-1334}}
\author[c]{Michael J. Randall\orcidlink{0009-0009-9806-2317}}
\author[c]{Joseph Seibert}
\author[b]{Grace Chesmore\orcidlink{https://orcid.org/0000-0001-6702-0450}}
\author[c]{Kevin T. Crowley\orcidlink{0000-0001-5068-1295}}
\author[d,e]{Nicholas Galitzki\orcidlink{0000-0001-7225-6679}}
\author[f,g]{Jon Gudmundsson\orcidlink{0000-0003-1760-0355}}
\author[h]{Kathleen Harrington\orcidlink{0000-0003-1248-9563}}
\author[i]{Bradley R. Johnson\orcidlink{0000-0002-6898-8938}}
\author[c]{JB Lloyd\orcidlink{0000-0003-1581-1626}}
\author[a]{Amber D. Miller}
\author[j]{Max Silva-Feaver\orcidlink{0000-0001-7480-4341}}
\affil[a]{University of Southern California, 825 Bloom Walk, Los Angeles, CA 90089, USA}
\affil[b]{University of Chicago, Address, 5720 South Ellis Avenue, Chicago, IL 60637, USA}
\affil[c]{University of California, San Diego, La Jolla, CA 92093, USA}
\affil[d]{Department of Physics, University of Texas at Austin, Austin, TX, 78712, USA}
\affil[e]{Weinberg Institute for Theoretical Physics, Texas Center for Cosmology and Astroparticle Physics, Austin, TX 78712, USA}
\affil[f]{Science Institute, University of Iceland, 107 Reykjavik, Iceland}
\affil[g]{The Oskar Klein Centre, Department of Physics, Stockholm University, AlbaNova, SE-10691 Stockholm, Sweden}
\affil[h]{Argonne National Laboratory, High Energy Physics Division. 9700 S Cass Ave, Lemont, IL 60439}
\affil[i]{Department of Astronomy, University of Virginia}
\affil[j]{Department of Physics, Yale University}

\authorinfo{Corresponding Author: Remington G. Gerras
\\E-mail: gerras@usc.edu}

\begin{document} 
\maketitle

\begin{abstract}
The Simons Observatory is a ground-based telescope array located at an elevation of 5200 meters, in the Atacama Desert in Chile, designed to measure the temperature and polarization of the cosmic microwave background. It comprises four telescopes: three 0.42-meter small aperture telescopes (SATs), focused on searching for primordial gravitational waves, and one 6-meter large aperture telescope, focused on studying small-scale perturbations. Each of the SATs will field over 12,000 TES bolometers, with two SATs sensitive to both 90 and 150 GHz frequency bands (SAT-MF1, and SAT-MF2), while the third SAT is sensitive to 220 and 280 GHz frequency bands. Prior to its deployment in 2023, the optical properties of SAT-MF1 were characterized in the laboratory. We report here on measurements of beam maps acquired using a thermal source on SAT-MF1, along with measurements of near-field beam maps using a holographic method that enables characterization of both the amplitude and phase of the beam response, yielding an estimate of the far-field radiation pattern received by the telescope. We find that the near-field half-width-half-maximum (HWHM) requirements are met across the focal plane array for the 90 GHz frequency band, and through most of the focal plane array for the 150 GHz frequency band. Namely, the mean of the bandpass averaged HWHM of the edge-detector universal focal plane modules match the simulated HWHM to 10.4 $\%$, with the discrepancy caused by fringing in the simulation. The measured radial profile of the beams matches simulations to within 2 dB from the beam center to at least the -10 dB level. Holography estimates of the far-field 90 GHz beams match the full-width-half-maximum from simulation within $1\%$, and the beam radial profiles deviate by less than 2 dB inside the central lobe. The success of the holography and thermal beam map experiments confirmed the optical performance were sufficient to meet the science requirements. SAT-MF1 was deployed to Chile in June, 2023. On-site observations are currently underway.
\end{abstract}


\section{INTRODUCTION}
\label{sec:intro}  
    The Simons Observatory (SO) is a ground-based telescope array dedicated to creating high-resolution and high-sensitivity temperature and polarization maps of the cosmic microwave background (CMB). It is currently deployed to the Atacama Desert and is composed of three small aperture telescopes (SATs) and one large aperture telescope (LAT). These telescopes work in tandem to observe the CMB over 6 frequency bands and over a large range of angular scales, using a combined $\approx$ 60,000 transition-edge sensor (TES) detectors \cite{SO_forecast_paper}. The six observed frequency bands are: the low frequency (LF) bands centered at 30 and 40 GHz, the middle frequency (MF) bands centered at 90 and 150 GHz, and the ultra high frequency (UHF) bands centered at 220 and 280 GHz. The LF bands are used to observe low-frequency synchotron emission, the UHF  bands are used to characterize galactic dust and foreground signal, while the MF bands are used to analyze the peak CMB signal. The LAT has a 6m aperture and an $8^{\circ}$ field of view, allowing for the LAT to resolve the CMB sky at arcminute angular resolution \cite{LAT_design}. The LAT's science goals focus on small-scale anisotropies. The SATs have a 42 cm aperture and an $35^{\circ}$ field of view. The SATs have an angular resolution of a half-degree, and will be used to study large-scale B-modes \cite{SAT_design}. Each SAT holds one optics tube and has $\approx$ 12,000 detectors. Two of them will be sensitive to MF (SAT-MF1 and SAT-MF2), and one will be sensitive to UHF (SAT-UFH). One of the primary science goals of SO is to observe or constrain the tensor-to-scalar ratio, \textit{r}, on the order of $r=0.01$ with uncertainty of $\sigma(r) < 0.003$ \cite{SO_white_paper}.
    
    SO requires accurate measurement of its beam shape and precise characterization of systematic errors throughout the optical system to observe the CMB sky at the precision needed to constrain \textit{r}. Intensity-only (thermal) beam maps are used in cosmology as confirmation and calibration of telescope optics, testing the angular response of the telescope as a thermal source scans across the field of view of the focal plane. The full-width-half-max (FWHM), ellipticity, and side-lobes of thermal beam maps provide an accurate characterization of the telescope beam shape \cite{beam_characteristics}, allowing for the deconvolution of the telescope beam from the observed CMB signal \cite{beam_leakage} \cite{Duivenvoorden_2019}.
    
    It is ideal to create beam maps in the far-field of the telescope before deploying the telescope to the site. However, observing far-field calibration sources from the laboratory is challenging. Near-field thermal beam maps act as a practical, first-order test of the telescope optics. Systematic errors in the observed near-field beam can be studied, allowing for problems to be rectified in the laboratory.
    
    Collaborators at the University of Chicago in 2021 applied the holographic method to measure the near-field beam intensity and polarization of the LAT-test receiver optics tube, providing an estimate of the far-field beam response of the telescope \cite{Chesmore_2022}, \cite{near-field_holog}. From late 2022 to early 2023, we performed near-field holography and thermal beam map experiments to study the telescope optics of SAT-MF1 at the University of California San Diego.  

    In this proceedings, we validate the near-field beam shape of the Simons Observatory SAT-MF1 via the near-field thermal beam maps, and estimate its far-field beam shape via the holography experiment. Section ~\ref{sec:methods} details the experimental setup for the holography and thermal beam map experiments, and outlines the analysis pipeline used to compare experimental results to simulations. Section ~\ref{sec:results} introduces the validation metrics of the optics, and discusses the results of the experiments compared against these metrics.

\section{METHODS}
\label{sec:methods}
    \subsection{Experimental Setup}
    In this section, we describe the experimental setup and data pipeline used to collect and analyze the near-field beam maps.

    \subsubsection{Near-Field Thermal Beam Experimental Setup}
    \begin{figure}[ht!]
        \centering
        \includegraphics[height=7cm]{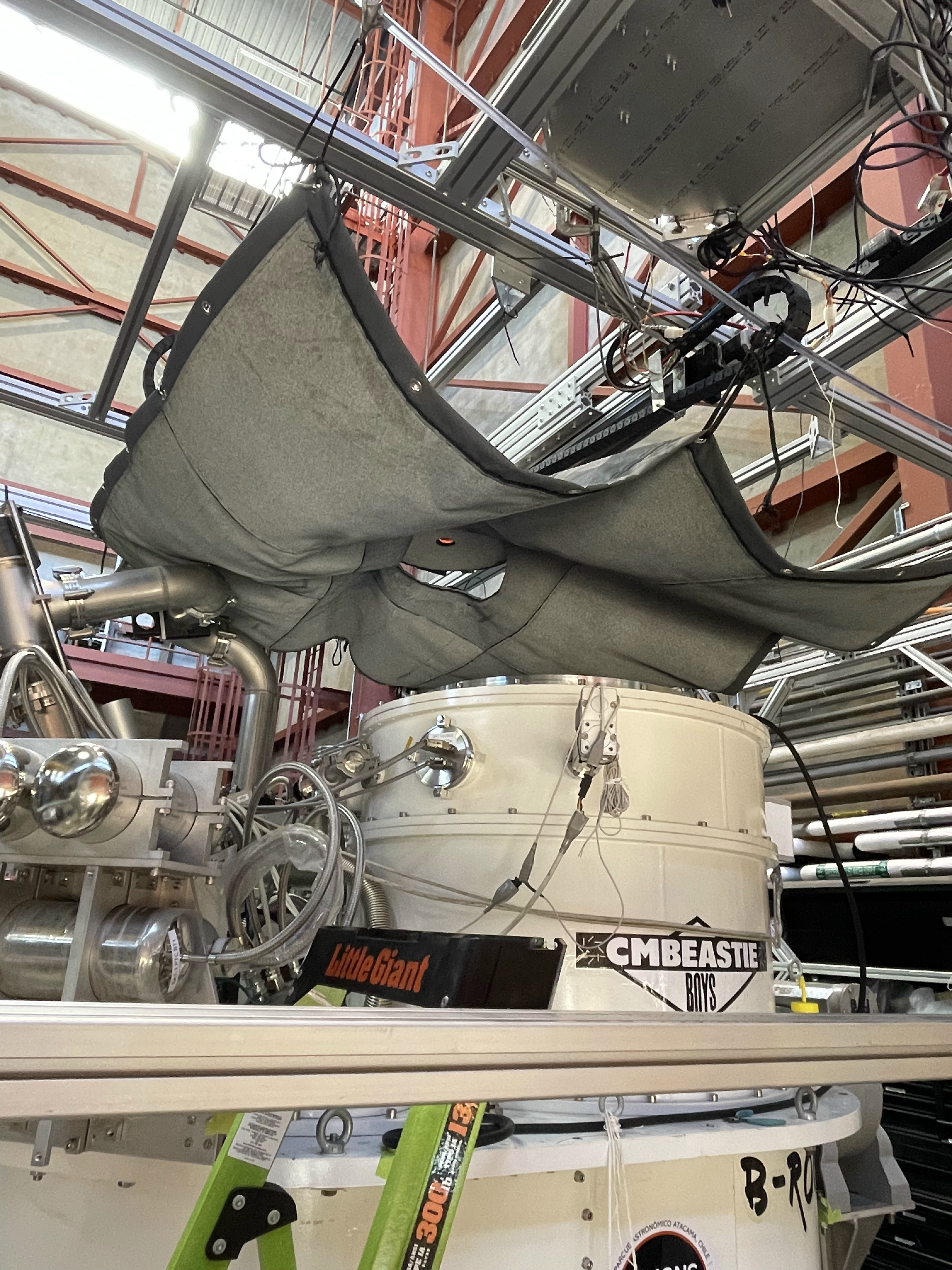}
        \caption{Experimental setup for acquiring thermal beam maps. The 8020 scaffolding supports the X/Y stages and the heat source and chopper plate above SAT-MF1. An Eccosorb blanket lies in the plane of the chopper plate to minimize reflections from electronics and the scaffolding. Note, SAT-MF1 is pointing at zenith.}
        \label{fig:thermal_setup}
    \end{figure}

    Design of the apparatus used to measure the near-field thermal beam maps of SAT-MF1 was guided by two primary considerations: the need to systematically scan a source of optical signal across the viewing window and continuously monitor the measurement status; and the need to measure the beam shape at a resolution comparable to simulation. 
    
    Figure~\ref{fig:thermal_setup} shows the experimental setup during collection of the beam maps. The thermal source was a heater attached to X/Y stages via a metal back-plate. The heater was a 6x6 cm ceramic Elstein SHTS/4 High Temperature Radiant Heater \footnote{\url{https://ihshotair.com/products/elstein-hts-panel-radiator-infrared-heater}} and comes with a thermocouple. The source was maintained at ~$760^\circ C$ and placed 81 cm above the aperture stop of SAT-MF1. The temperature of the thermocouple signal was readout with a Labjack T7 \footnote{\url{https://labjack.com/products/labjack-t7}}. The X/Y stages are constrained to move in two dimensions and were held safely above SAT-MF1 by being bolted onto an Aluminum 8020 support structure. The beam source was moved in a 76x76 cm square with 1 cm resolution via raster scan, sufficient to measure at least twice the expected beam size. The resolution of the experimental beam map was chosen to match the resolution of the simulation. Commanding and monitoring the position of the X/Y stages, and monitoring the temperature of the heat source, were all done via the Simons Observatory Control System (SOCS \cite{socs}). Current was sent through the source using a variac set by hand.

    Reflections from the 8020 scaffolding and motors were mitigated by hanging a blanket lined with Eccosorb AN-72\footnote{\url{https://www.laird.com/products/absorbers/microwave-absorbing-foams/multi-layer-foams/eccosorb-an}} at $\approx 300$ K in the field-of-view of SAT-MF1, with the beam source visible to the detectors through a 3.8 cm aperture. To address long-timescale drifts, the source is chopped at 6 Hz using spinning fan blades covered in Eccosorb AN-72. The frequency of the dual-blade fan was monitored via an oscilloscope kept near SAT-MF1.

    To prevent the detectors from saturating, neutral density filters (NDFs) were installed onto the focal plane of SAT-MF1\cite{SAT1_optical_testing}. We blocked off a fraction of our detectors by covering them with a copper rhombus to investigate electrical cross-talk and magnetic pickup.      
    
    \subsubsection{Near-Field Holography Experimental Setup}
    \begin{figure}[ht!]
        \centering
        \includegraphics[height=7cm]{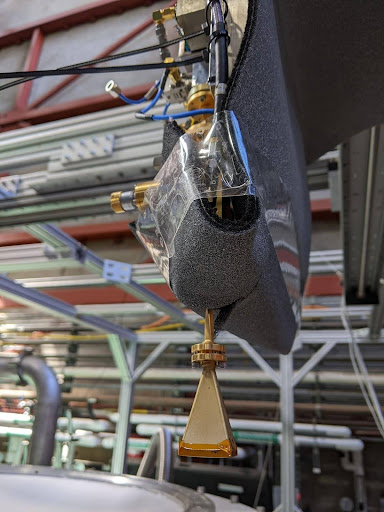}
        \includegraphics[height=7cm]{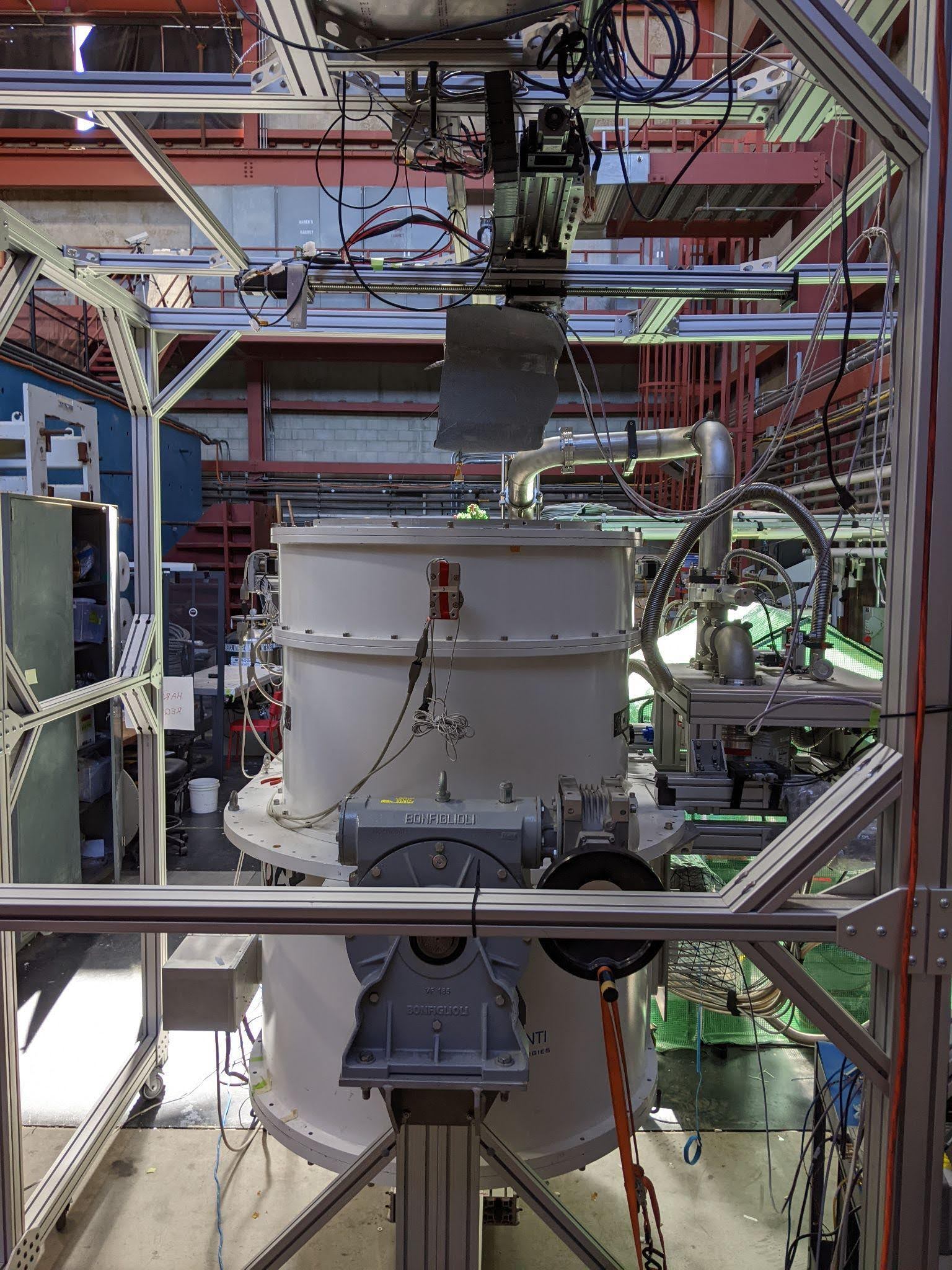}
        \caption{Holography experimental setup. Left: Holography frequency source with waveguide. Right: Holography setup above the SAT-MF1 window with Eccosorb covering.}
        \label{fig:holography_source}
    \end{figure}

    
    The near-field holography source emits a monochromatic signal tuned to frequencies within the 90 GHz passband ([74, 112] GHz) and 150 GHz passband ([124, 174] GHz). We tuned the holography source to emit both co-polar and cross-polar signal at 5 GHz intervals over the frequency range of [85, 135] GHz by using a waveguide twist to rotate the polarization of the signal by $90^{\circ}$. Scheduling constraints limited the number of single frequency beam maps tested, allowing only for frequencies within the 90 GHz passband to be used in the analysis of this proceedings.      

    Section 3 of Chesmore et al \cite{Chesmore_2022} provides a schematic diagram of the holography source and receiver used for tuning. Similar to the thermal beam experimental setup, the holography source was also suspended above SAT-MF1 using the 8020 structure. We used Eccosorb AN-72 to absorb reflections and radio signal from the holography electronics as shown in figure ~\ref{fig:holography_source}, and scanned the holography source across the field-of-view of SAT-MF1 using the same X/Y stages and motor controllers. 
    
    There are three characteristics of the holography experimental setup that are distinct from the thermal beam map experimental setup. The holography signal is detected by a special purpose receiver located at the radial edge of the focal plane (16.3 cm from center), depicted as stars in the left panel of figure \ref{fig:sat_fpa}. Second, the holography receiver is not at risk of saturating from the 300 K background signal of the lab, so does not require the NDF.  Lastly, the holography experiment requires the SAT-MF1 focal plane to be cooled to 4 K, rather than the operating temperature of 100 mK used for the thermal beam maps. This is because SAT-MF1 must be warm enough for the holography receivers to operate, and cold enough for the 40 K and 4 K alumina filters to be transmissive \cite{SAT_design}. 
    \subsection{Analysis Pipeline}
    In this section, we detail the analysis pipeline and the methods used to analyze near-field thermal and holography beams. First, we discuss the simulation software used to simulate the SAT-MF1 near-field thermal beams, Ticra Tools\footnote{\url{https://www.ticra.com/}}, to which the measured beams are to be compared. Then we discuss the analysis methods used to analyze both types of beam map.

\begin{figure}[ht!]
        \centering
        \includegraphics[scale=1.81]{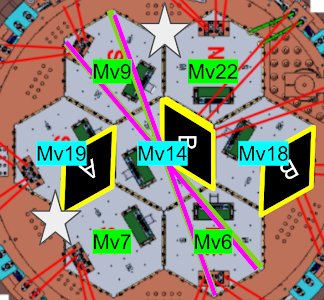}
        \includegraphics[scale=0.45]{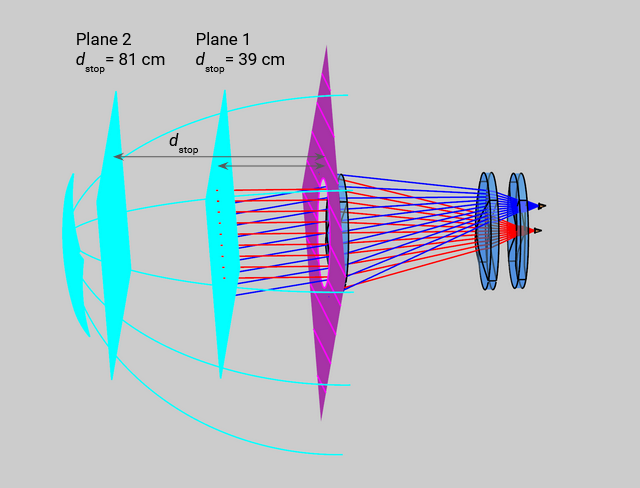}
        \caption{Left: Focal Plane of SAT-MF1. Green and blue texts provide the names of the Universal Focal-plane Modules (UFMs).  Black rhombuses depict blocked off detectors via copper rhombus attached to UFM. Stars show the locations of holography readout sensors. Right: \textbf{Ticra Tools} model and simulation of the thermal beam and holography setup using SAT-MF1 optics. The holography source was 39 cm from the aperture stop of SAT-MF1, and the thermal beam source was 81 cm from the aperture stop of SAT-MF1. Observing pixels at the center and radial edge (16.3 cm) on the focal plane are depicted on the right side, with red and blue light line terminating at these pixels. Pink plane: the aperture stop of SAT-MF1. Light blue planes: Planes from which beam sources are emitted.}
        \label{fig:sat_fpa}
    \end{figure}

   Using Ticra Tools, we simulate the SAT-MF1 optical response to the near-field thermal source by calculating the electric field through each optical interface. Fig~\ref{fig:sat_fpa}, right, depicts the simulation setup along with the optical components in SAT-MF1. The optics contain three refractive lenses, an aperture stop, feedhorns, and the focal plane. This simulation includes two planes of emission: 81 cm from the aperture stop, corresponding to the near-field thermal beams; and 39 cm from the aperture stop, corresponding to the holography near-field beam maps. Beam maps are simulated at two positions in the focal plane: one at the center of the focal plane; and the other at the edge of the focal plane, at a radial distance of 16.3 cm from the center. Ticra Tool simulations calculate beam radial profiles from monochromatic sources, so we average simulated beam maps across observing frequency bands to compare with the observed thermal beam maps. The simulated beam radial profiles are presented in section \ref{sec:results}. We weight the intensity in each simulated beam map by the bandpass transmissivity of the detectors inside SAT-MF1\cite{SAT1_optical_testing}. The cryogenic half-wave plate\cite{Yamada_2024} was installed in SAT-MF1 for both the holography and thermal beam map measurements, however it was not rotating during the measurements.

    \subsubsection{Thermal Beam Analysis Pipeline}
    We discuss the pre-processing pipeline that is used to produce thermal beam maps from time-ordered-data (TOD). Then, we describe the validation metrics used to compare the simulated beam radial profiles to the observed SNR-weighted average beam radial profiles. 
   
    The first step in the analysis pipeline is to pre-process each of the detector TODs using the sotodlib library\footnote{\url{https://github.com/simonsobs/sotodlib}}. We first linearily detrend the timestream data and pass it through a 4th order high-pass Butterworth filter with 2Hz cutoff frequency. This process removes trending signal in the detectors, and filters out low-frequency background noise from the laboratory. The TODs are then demodulated at the chop frequency (6 Hz), the X/Y stage positions are mapped to the timestamps in the the detector timestreams, and the amplitude of the detected signal at each pixel in the beam map is computed as the average value of the TOD over the 0.5s integration time, the duration in which the thermal source remains still. We then fit for the centroid of the observed beam maps for the 90 GHz and 150 GHz detectors using two different methods.

\begin{figure}[ht!]
        \centering
        \includegraphics[width=7cm, height=7cm]{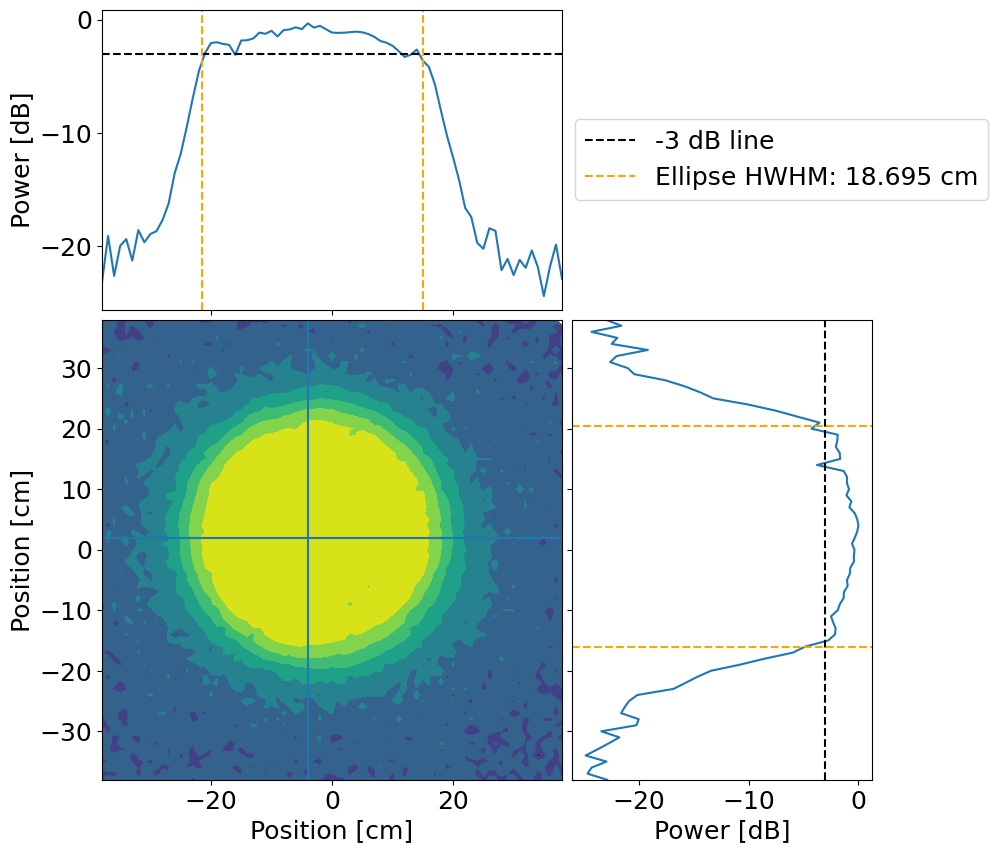}
        \includegraphics[width=8cm, height=7cm]{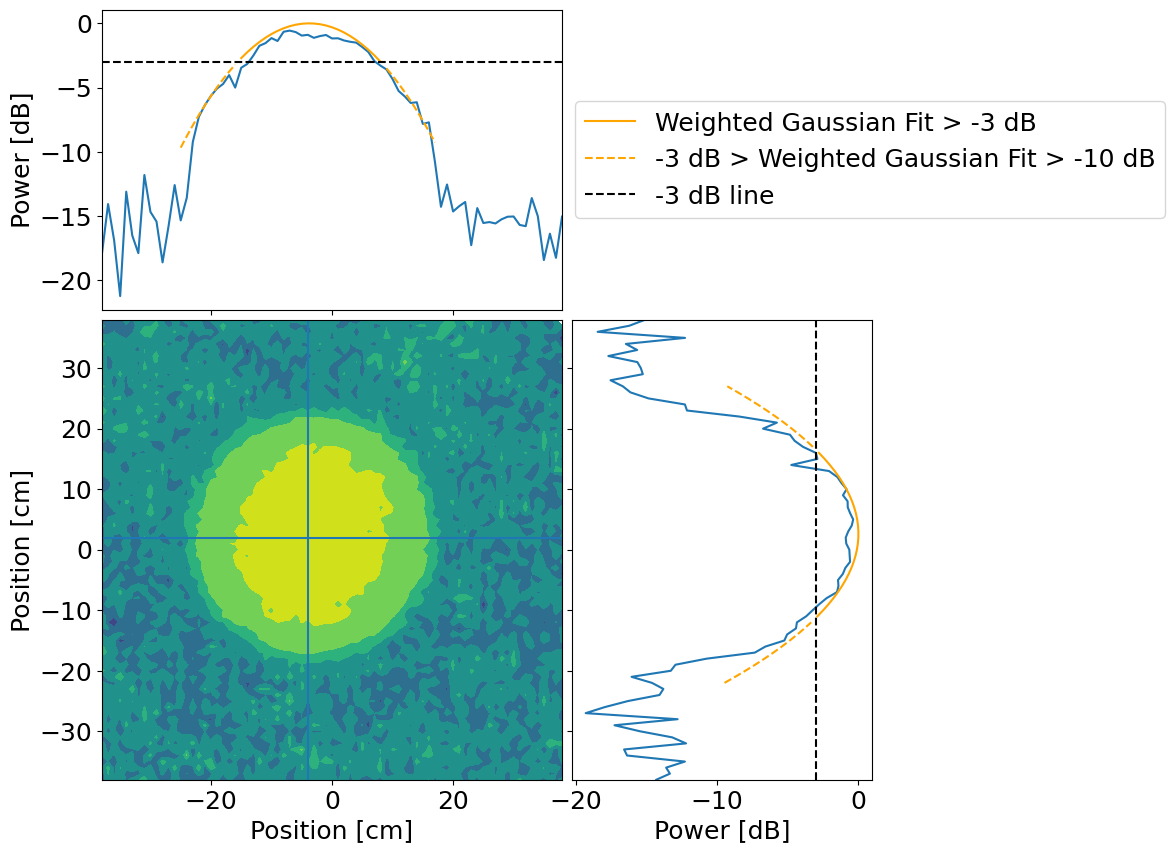}
        \caption{Observed thermal beam maps after data processing. Left: 90 GHz beam on a central pixel with power slices through the center of the beam. Orange: HWHM fit from an ellipse. Black: -3 dB line. Right: 150 GHz beam on a central pixel with power slices through the center of the beam. Orange: Weighted Gaussian fit. The fit is represented by a line where its intensity is $> -3$ dB and thus larger weights, and by a dashed orange line where its intensity is between -3 and -10 dB.}
        \label{fig:central_thermal_beams}
    \end{figure}

    First, we discuss fitting the centroid of the observed beams for detectors within the 150 GHz passband. The observed beams  are non-Gaussian below -5dB due to the aperture stop cutting off signal past 21 cm from the center of the beam, an example can be seen on the right panel of figure ~\ref{fig:central_thermal_beams}. To account for this, we implement a weighted fitting of a 2D Gaussian 
    \begin{equation}
        A\exp{(-a(x-x_{0})^2 + 2b(x-x_{0})(y-y_{0}) + c(y-y_{0}})^2)
    \end{equation}
    with 
    \begin{equation}
        a = \dfrac{\cos{\theta}^2}{2\sigma_{x}^2}  + \dfrac{\sin{\theta}^2}{2\sigma_{y}^2}, 
        b = \dfrac{-\sin{\theta}\cos{\theta}}{2\sigma_{x}^2}  + \dfrac{\sin{\theta}\cos{\theta}}{2\sigma_{y}^2}, 
        c = \dfrac{\sin{\theta}^2}{2\sigma_{x}^2}  + \dfrac{\cos{\theta}^2}{2\sigma_{y}^2}
    \end{equation} 
    to the observed beam map. A is the amplitude, $\theta$ is the angle of rotation, $x_{0}$ and $y_{0}$ are the center in centimeters, and $\sigma_{x}$ and $\sigma_{y}$ are the spread of the observed 2D Gaussian. The weights are the inverse of the observed peak-normalized and logarithmic intensity. We then use the fitted centroid as the center of the observed beam in the 150 GHz passband detectors.

    Next we discuss fitting detectors in the 90 GHz passband. These detectors did not observe Gaussian near-field beams because the 90 GHz detectors were saturated in addition to the 42 cm aperture stop cutting off lower power signal. This produces a relatively flat beam with sharp cutoffs 21 cm away from the beam center (left panel of figure ~\ref{fig:central_thermal_beams}). To account for the flat-top shape of the 90 GHz beams, we implement an ellipse fitting and re-fitting algorithm to find the beam centroid. The first fitted ellipse estimates the beam centroid from the brightest pixel in the observed beam. This first pass calculation may under-report the beam half-width-half-max (HWHM) by centering the ellipse around the brightest pixel, rather than the geometric center of the beam. Thus, we smooth-out the resulting beam center by averaging the observed power around the center pixel within a 7.5 cm radius circle, and re-fit the ellipse to calculate the reported beam centroid. Details can be found in Appendix \ref{sec:appen_a}.

    We calculate composite beam maps from an SNR-weighted average of all detectors within the 90 and 150 GHz frequency passbands within a UFM. Comparisons between observed beam maps and simulated beam maps are done using the radial profiles of composite beams as done in the BICEP-Keck collaboration \cite{BK11}. Composite beams provide a measure of the optical performance across the focal plane array (FPA) and beam radial profiles provide a quantitative description of the radial power distribution of the beam. The metrics of comparison are the half-width half-max (HWHM), and the deviation in beam power up to -10dB of the simulated beam. Detectors that were blocked off by a copper rhombus or failed to be accurately fit did not contribute to the composite beam map. Failed fits were defined as observing the beam center outside the perimeter of the beam map, or having a HWHM  larger than the radius of the aperture stop.
    
    To make comparisons to the Ticra Tools simulation, we created composite beam maps in the 90 and 150 GHz frequency passbands using UFMs at the center of the FPA (Mv14), and averaging over all UFMs at the edge of the FPA (Mv22, Mv18, Mv6, Mv19, Mv9) excluding Mv7. UFM Mv7 was not able to be read out due to readout wiring faults that have since been corrected.
    
    \subsubsection{Holography Beam Analysis Pipeline}
     We use the near-field phase and amplitude measurements of the holography experiment to estimate the far-field response of SAT-MF1. This estimate is performed using the antenna-receiver relationship: the electric field measured at the aperture is the Fourier Transform of the far-field radiation pattern \cite{near-field_holog}:

    \begin{equation}
        B(\theta_{x},\theta_{y}) = \int_{aperture} E(x,y)\exp{(i \dfrac{2 \pi}{\lambda} (\theta_{x}x + \theta_{y}y))} dxdy
    \end{equation}
    Integrating over the area of the aperture, E(x,y) is the complex electric field measured at the aperture, $\lambda$ is the wavelength, and B($\theta_{x}$,$\theta_{y}$) is the estimated far-field radiation pattern. 

    We measured both the co-and cross-polar amplitude and phase of the near-field holography beam maps. Then we band-averaged the co-polar 90 GHz frequency band [85-110] GHz measurement to project its far-field response. The far-field beam response estimate is fit to a 2D Gaussian to calculate the centroid of the beam, then we calculate its beam radial profile to compare the full-width-half-max (FWHM) and beam shape against simulation. We used a single frequency (90 GHz), far-field beam map simulation\cite{Dachlythra_2024} to compare with the holography far field estimate.
    
    Comparing the band-averaged far-field estimate from holography against the single frequency simulation at the approximate median of the passband, 90 GHz, is an imperfect comparison, but is sufficient for the purposes of a pre-deployment validation of the optical performance. Full characterization of the SAT-MF1 far-field beam shape will be done on-site during the commissioning phase of the telescope.

\section{RESULTS}
\label{sec:results}
In this section, we discuss the near-field beam properties as measured by the thermal beam map measurement, and the far-field radiation response projection from the holography measurement. We use the thermal beam map measurement to validate the near-field beam properties of SAT-MF1 and the holography measurement to project SAT-MF1's far-field radiation response.
\subsection{Thermal Beam Results}
\begin{figure}[ht!]
        \centering
        \includegraphics[height=7cm]{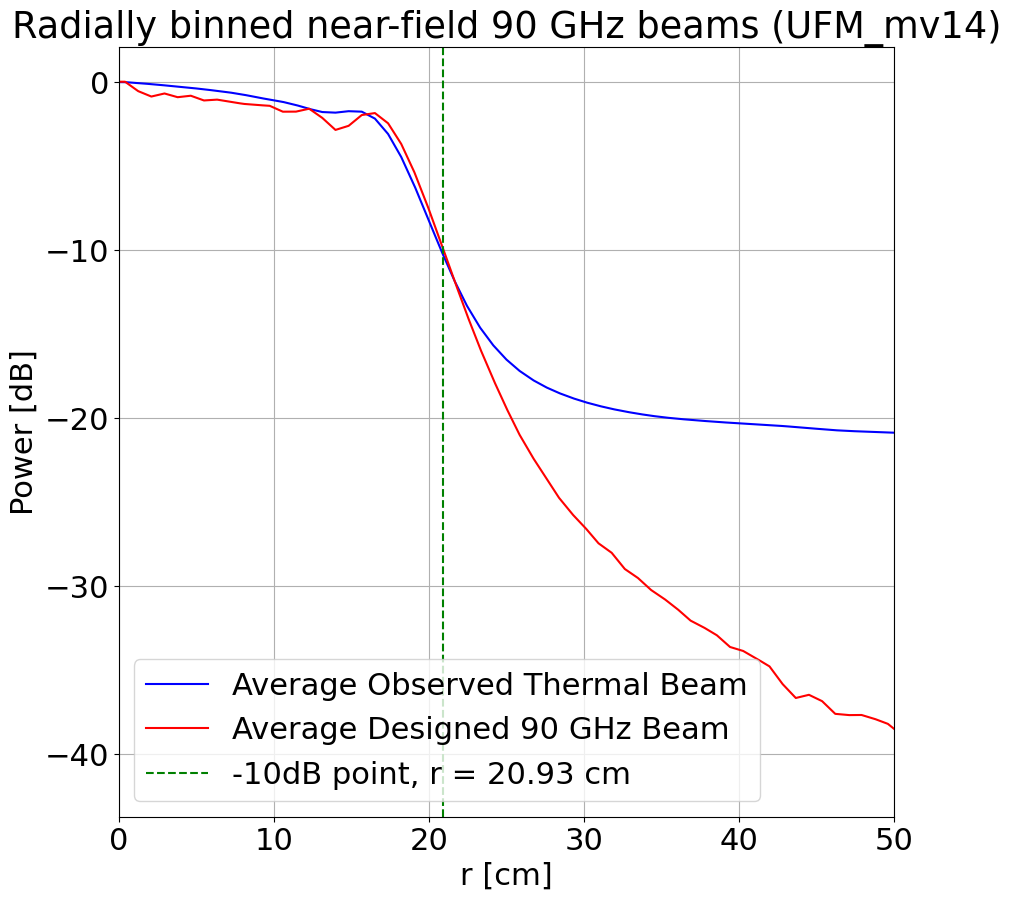}
        \includegraphics[height=7cm]{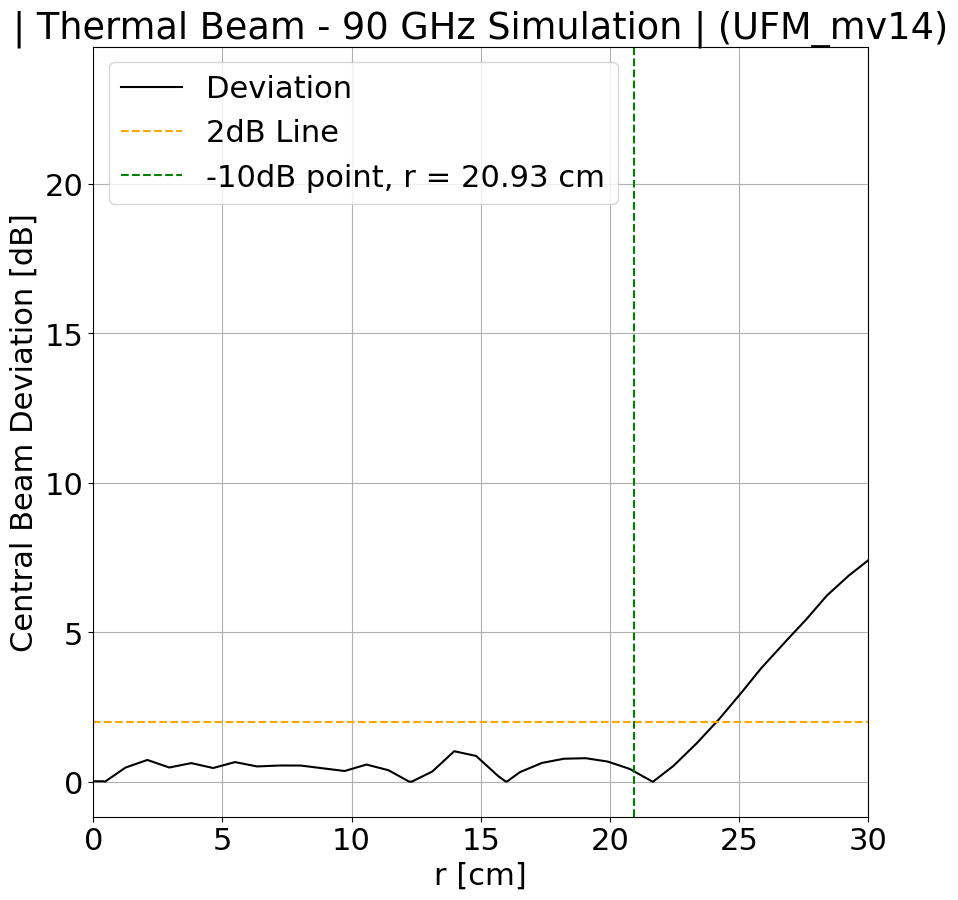}
        \includegraphics[height=7cm]{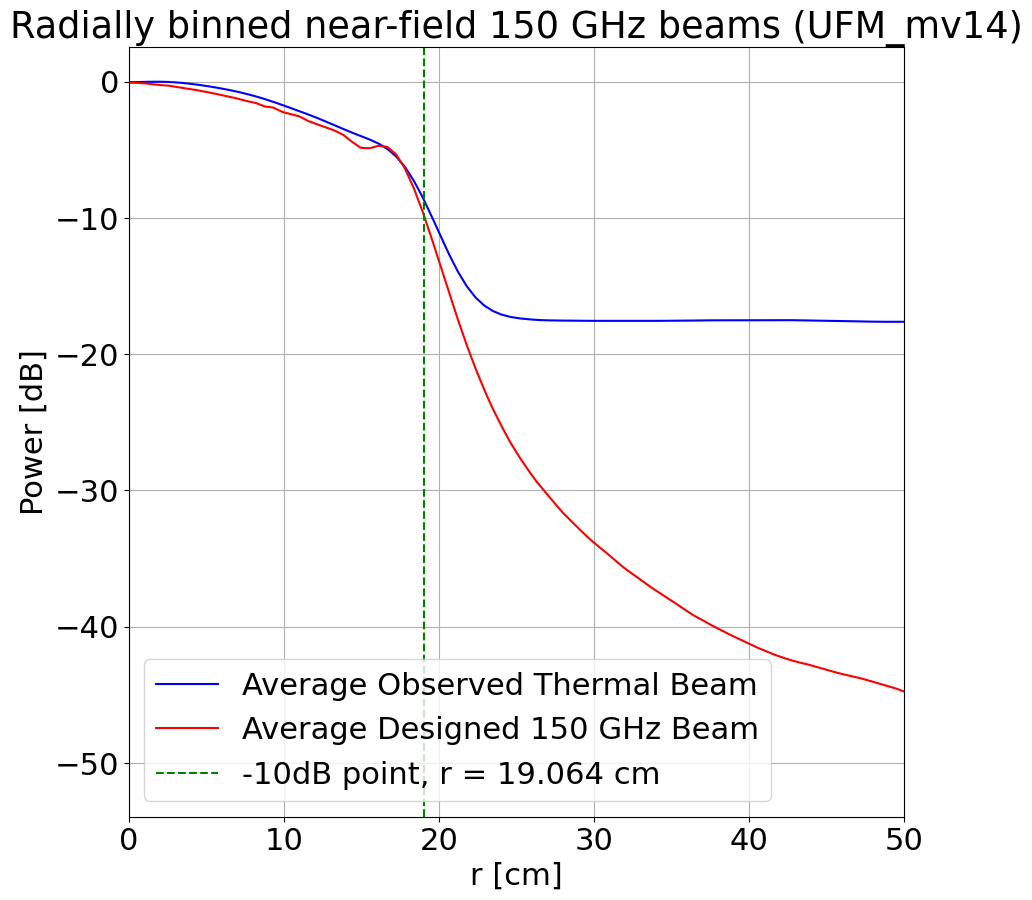}
        \includegraphics[height=7cm]{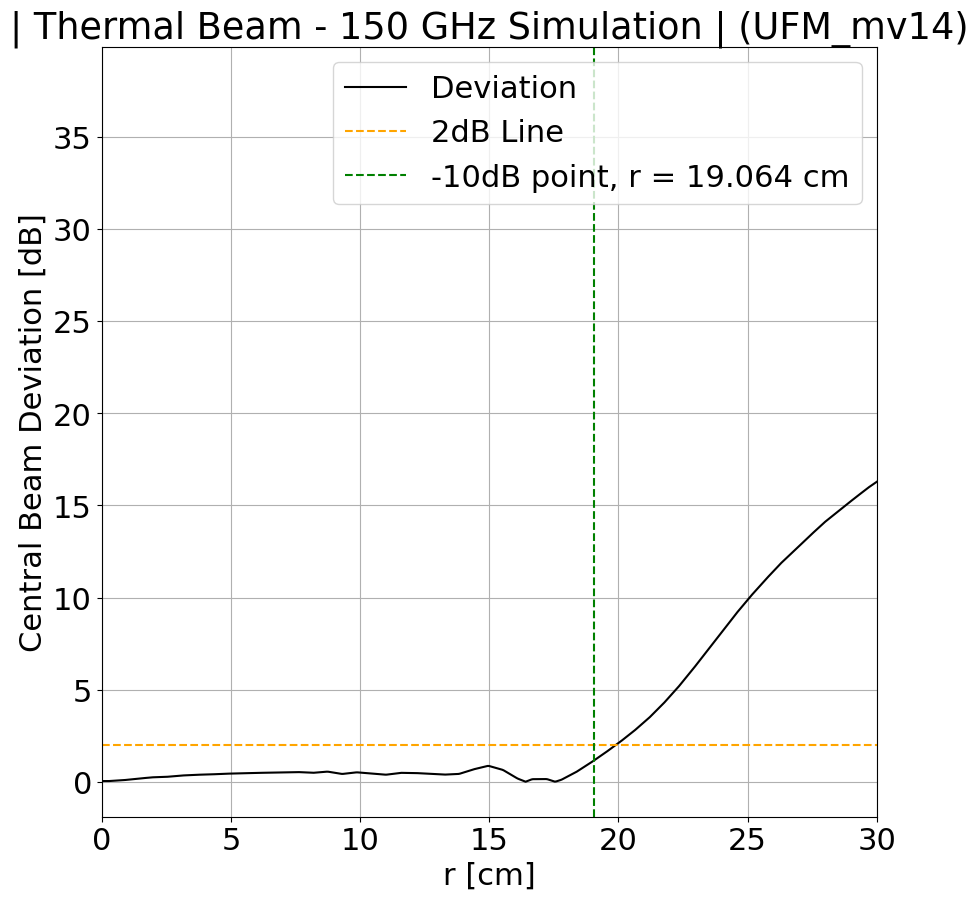}
    
        \caption{Near-field thermal beam profiles for the central UFM (Mv14). Top: 90 GHz beam analysis. Left panel is the radial beam profile of the composite beam (blue) versus the simulated beam profile (red). Right panel is the beam power deviation within the central beam. The beam deviation must be $<$ 2 dB (dotted orange line) up to - 10 dB of simulation (dotted green line) to meet science requirements. Bottom: 150 GHz beam analysis. Left panel is the observed thermal beam (blue) versus the simulated band pass averaged beam (red). Right panel is the beam power deviation within -10 dB of the simulated beam.}
        \label{fig:mv14_near-field_profile}
    \end{figure}

\begin{figure}[ht!]
        \centering
        \includegraphics[height=7cm]{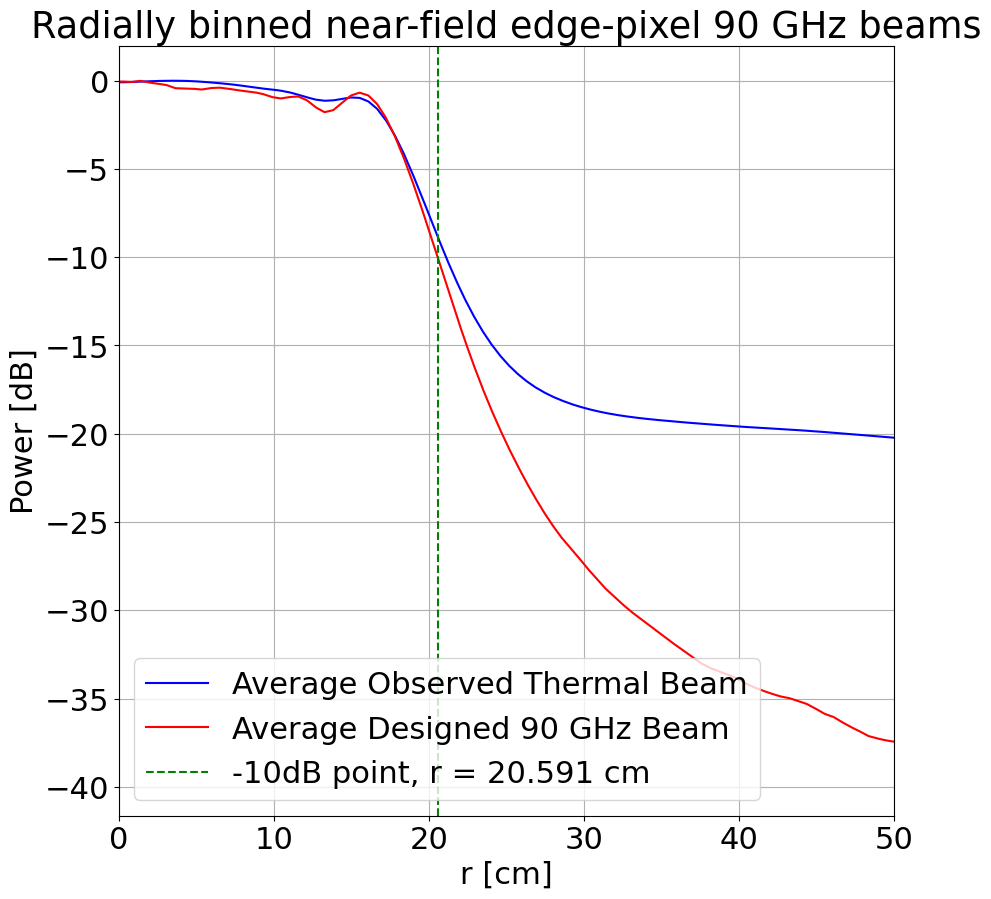}
        \includegraphics[height=7cm]{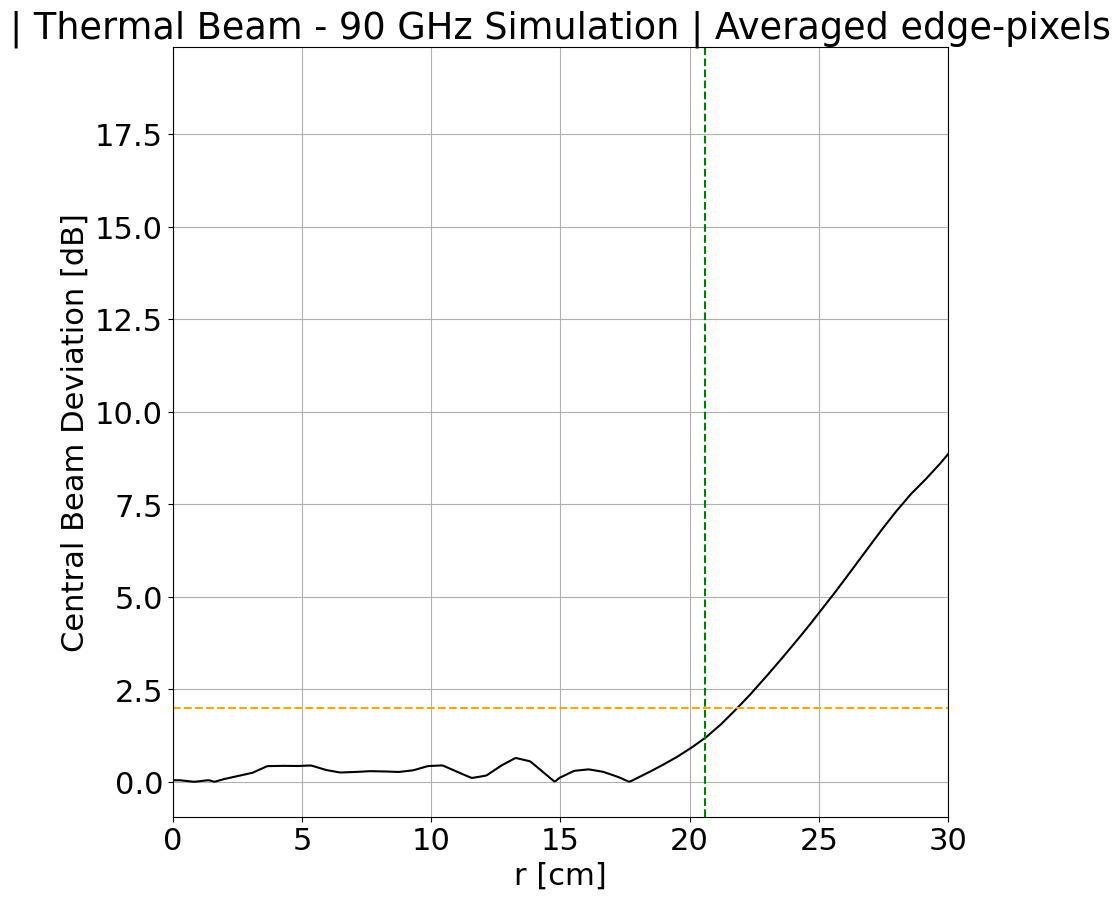}
        \includegraphics[height=7cm]{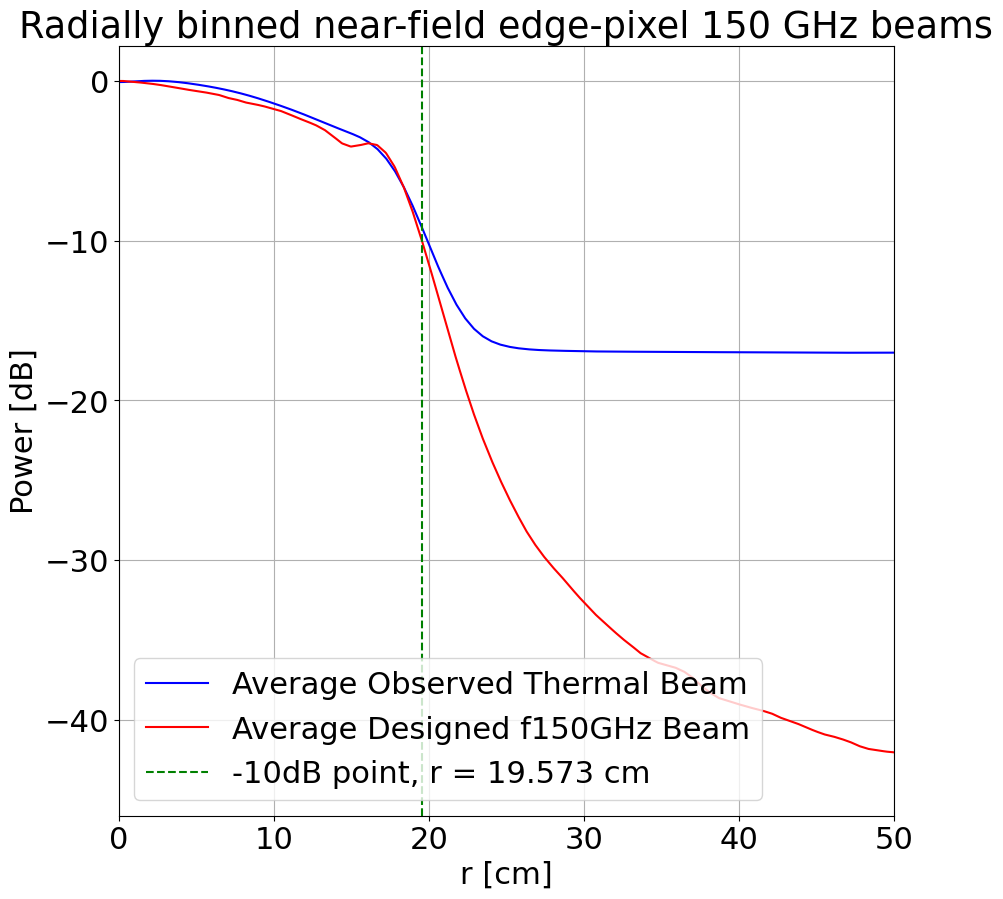}
        \includegraphics[height=7cm]{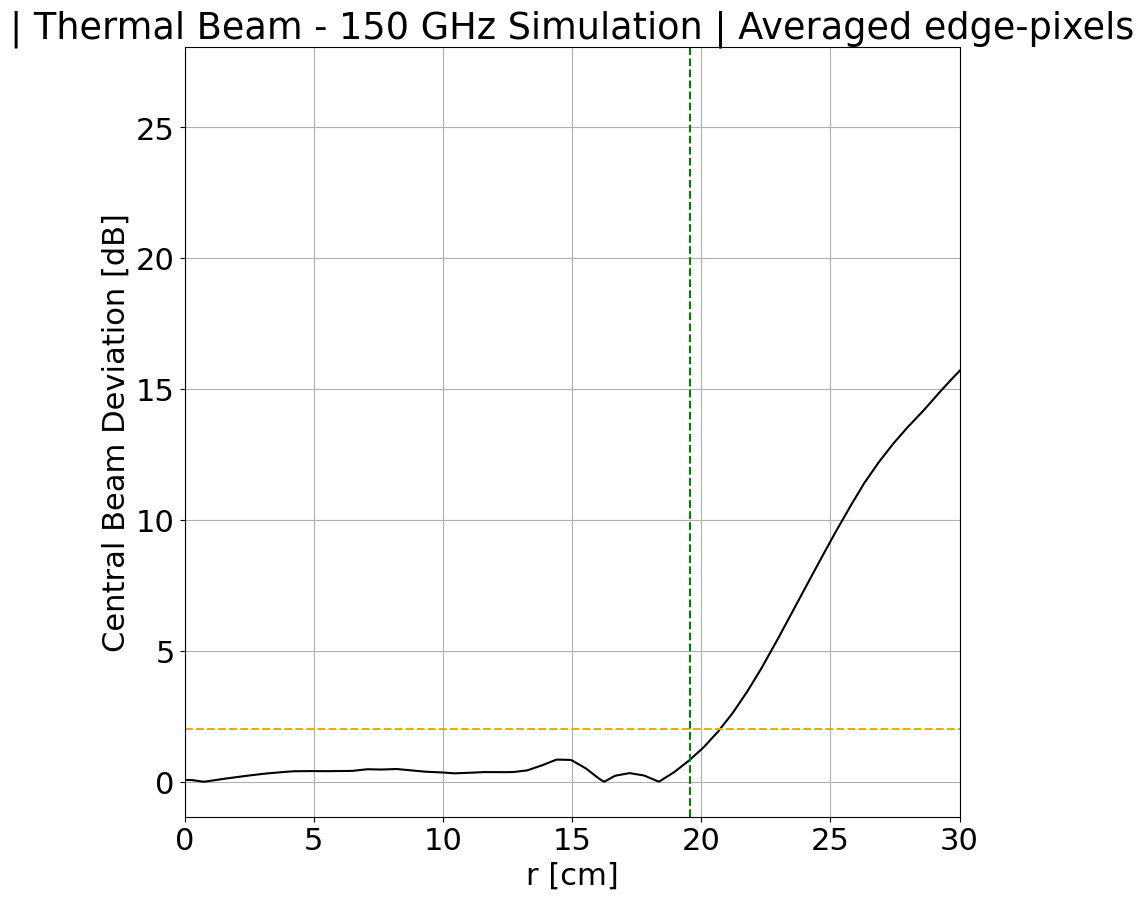}
    
        \caption{Averaged near-field thermal beam profiles for UFMs on the radial edge of the FPA. Top: 90 GHz beam analysis. Left panel is the radial beam profile of the composite beam (blue) versus the simulated beam profile (red). Right panel is the beam power deviation within the central beam. Bottom: 150 GHz beam analysis. Left panel is the observed thermal beam (blue) versus the simulated band pass averaged beam (red). Right panel is the beam power deviation within -10 dB of the simulated beam.}
        \label{fig:mv22_near-field_profile}
    \end{figure}

    Figures ~\ref{fig:mv14_near-field_profile} and ~\ref{fig:mv22_near-field_profile} show results of the comparison between the thermal beam measurements and the simulations for both the 90 and 150 GHz passbands at the center and edge of the focal plane, respectively. The observed near-field HWHM, defined as the radial distance from the center of the beam to the point the beam crosses -3 dB in power, must be within 10 $\%$ of the simulated beam HWHM, and the deviation in beam power must not be greater than 2 dB up to the -10 dB point in the simulated beam. The center UFM beam deviations of both 90 and 150 GHz passbands remain  below 2 dB up to -10 dB of the simulated beam. The deviation between observed beams and simulated beams at the edge of the FPA remain within specifications as well. The simulated beams experienced more fringing than the observed beams because the simulation calculated single-frequency near-field radiation patterns across each passband, while the observed thermal beams are band-averaged across the entire 90 and 150 GHz passbands. The monotonically increasing deviation between observed and simulated beam radial profiles at the -10 dB to -15 dB level is due to measurement noise.
    
\FloatBarrier

\FloatBarrier

    The HWHMs from the observed and simulated thermal beams at the center (Mv14) and averaged around the edge of the FPA, along with the averaged difference (Error) between the two, are listed in table ~\ref{tab:hwhm_comparisons}. The uncertainty in the HWHM is the standard deviation of the observed and simulated near-field beams. All HWHMs have an averaged difference within 10 $\%$, except for the 150 GHz edge-UFMs. The simulated HWHM value for the 150 GHz edge-UFMs is made smaller by fringing in the simulation.
        
    

    \begin{table}[ht!]
        \centering
        \begin{tabular}{|c|c|c|c|}
        \hline
             Parameter & Design Value (cm) & Measured Value (cm) & Error ($\%$) \\
             \hline
             Central HWHM 90 GHz & $17.9 \pm 0.1$& $17.5 \pm 0.4$ & 2.2 \\
             Edge HWHM 90 GHz & $17.9 \pm 0.1$ & $17.8 \pm 0.5$ & 0.6 \\
             Central HWHM 150 GHz & $12.6 \pm 0.1$ & $13.5 \pm 0.5$ & 7.1 \\
             Edge HWHM 150 GHz & $13.5 \pm 0.1$ & $14.9 \pm 0.9$ & 10.4 \\
        \hline
        \end{tabular}
        \caption{Table of simulated HWHM and observed HWHM for central and edge pixels of 90 and 150 GHz passbands}
        \label{tab:hwhm_comparisons}
    \end{table}

    \subsection{Holography Results}
    The near-field intensity beam maps measured from holography are cross-checked by the near-field thermal beam maps and near-field simulations. The holography near-field intensity beams are validated by passing the same requirements as the thermal beams: the HWHM must match simulation to within 10 \%, and the beam deviation must be $<$ 2 dB up to -10 dB of the simulated beam. Upon validation, 
    we used holography to measure three SAT performance criteria: the 90 GHz far-field response has full-width-half-max (FWHM) of $< 30$ arcmin, the integrated power within the central lobe is $> 98\%$, and lastly, the central lobe deviation is less than 2 dB between the simulated and holography projected far-field beams. The central lobe width is defined as the radial distance from the center of the simulated beam to its first airy null, a distance of 30 arcmin. The simulations and holography results are expected to agree within 2 dB from the center of the beam up to 30 ". We expect deviation beyond 30 " because taking the Fast-Fourier Transform of a noisy near-field radiation pattern would smear the first airy null across the rest of the data. 
    
\begin{figure}[ht!]
        \centering
        \includegraphics[scale=0.6]{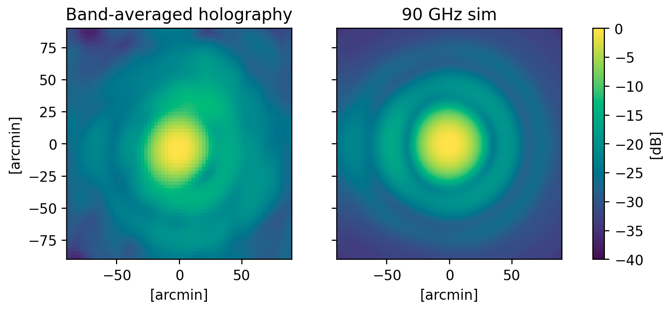}
        \caption{Band-averaged 90 GHz far-field projection from holography measurements (left) and single frequency 90 GHz simulation (right).}
        \label{fig:farfield_prof}
    \end{figure}

\begin{figure}[ht!]
        \centering
        \includegraphics[width=10cm, height=6cm]{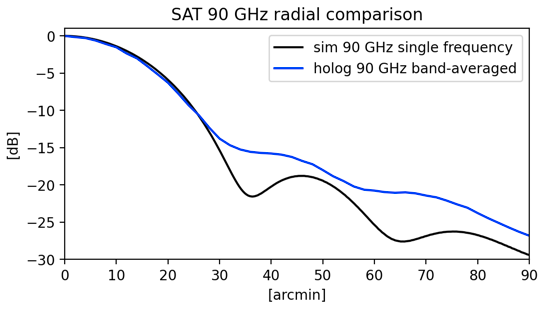}
        \includegraphics[width=10cm, height=6cm]{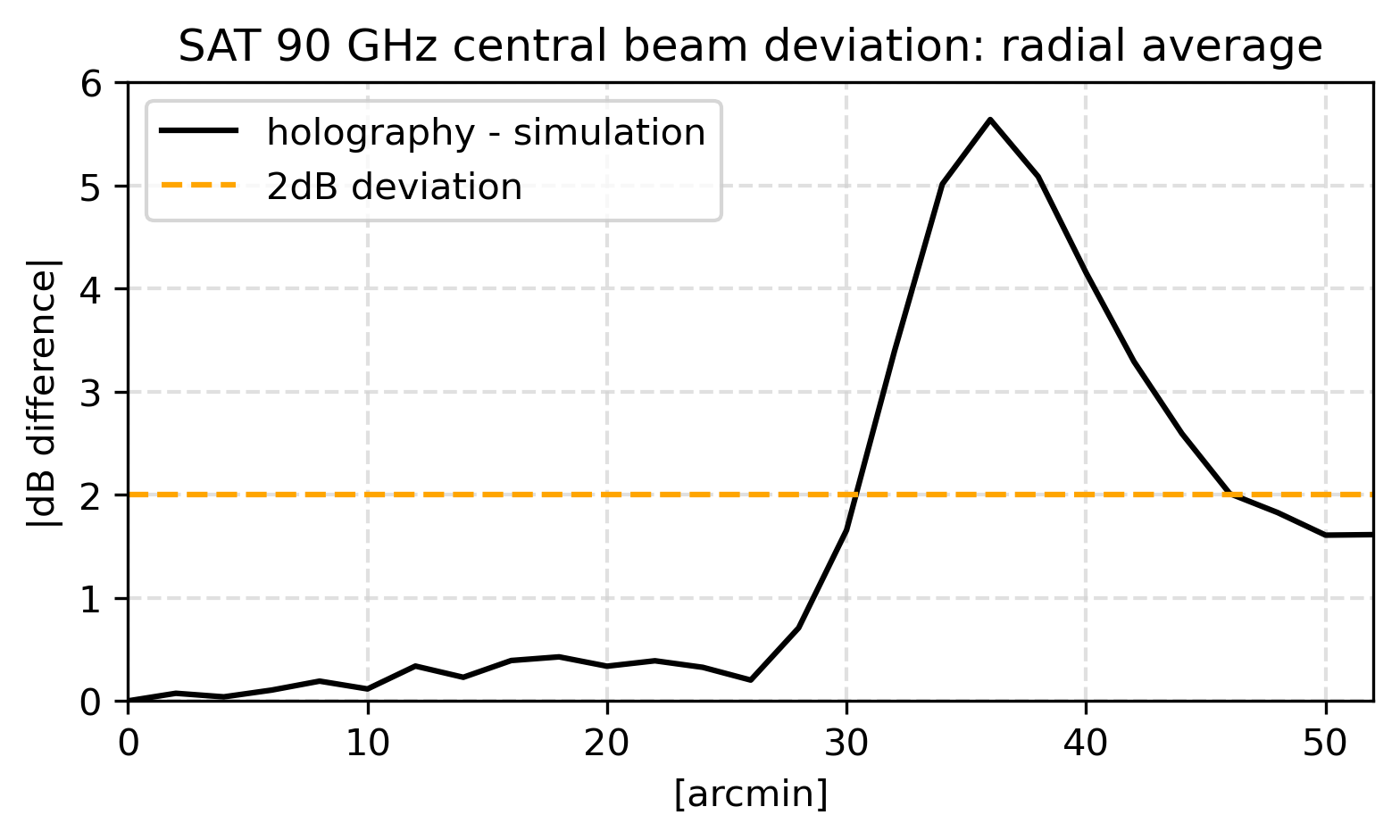}
        \caption{Far-field beam radial profiles from the holography projection and 90 GHz simulation. Top: Band-averaged 90 GHz holography projection (blue) and the simulated single frequency 90 GHz simulation (black) to 90 arcmin from the beam center. Bottom: Beam deviation of the holography projection and the 90 GHz simulation. The x-axis goes up to a radial distance of $\approx50$ arcmin from the beam center.}
        \label{fig:far-field-comps}
    \end{figure}

    Figure \ref{fig:farfield_prof} compares the 90 GHz band-averaged far-field beam response projected from holography measurements (85-115 GHz) to the simulated far-field beam response of a single frequency beam at 90 GHz. Figure ~\ref{fig:far-field-comps} compares the beam radial profile of the 90 GHz far-field response projected by holography, and the single 90 GHz simulation. The FWHM from the holography projection is $27.96$ arcmin while the FWHM from the single frequency simulation is $27.89$ arcmin, a deviation of $0.25 \%$. The beam power deviation stays within 2 dB everywhere except at the airy null of the simulation, $\approx$ 30 arcmin from the beam center. The integrated power measured within the central lobe is $97.4 \%$, slightly below the stated performance criteria. We believe the near-field scattering is intrinsic to the holography setup, rather than the telescope optics, is likely to be responsible for this small discrepancy.


\FloatBarrier

\FloatBarrier

\section{CONCLUSIONS}
\label{sec:conclusions}
We analyzed the measurements of the near-field thermal and holography beams on central and edge-pixels of the FPA for the Simons Observatory's SAT-MF1. Results from the near-field thermal beams confirm the designed beam shape to within 3 $\%$ for the 90 GHz frequency band, and within 10 $\%$ for the 150 GHz frequency band except for the edge-UFMs. This discrepancy is caused by fringing in the simulated HWHM.  The near-field holography results provide far-field projections that satisfy requirements from the simulated far-field beam within the central lobe. The slightly elevated side-lobe power relative to simulations in the holography estimated beam are likely caused by near-field scattering in the holography experiment. This is to be confirmed with commissioning measurements taken throughout 2024. The success of the holography and thermal beam map experiments confirmed the optical performance is sufficient to meet the science requirements, allowing for SO to deploy SAT-MF1 in June 2023 with first light achieved in October 2023. Characterization of SAT-MF1 optics are an important area of study for SO, with commissioning data currently ongoing.

\appendix    

\section{ALGORITHMS}
\label{sec:appen_a}
\subsection{Ellipse Fitting and Re-fitting Algorithm}
    This Appendix details the ellipse fitting and ellipse re-fitting algorithm used to calculate the beam centroid observed by the 90 GHz detectors.

\begin{figure}[ht!]
        \centering
        \includegraphics[scale=0.35]{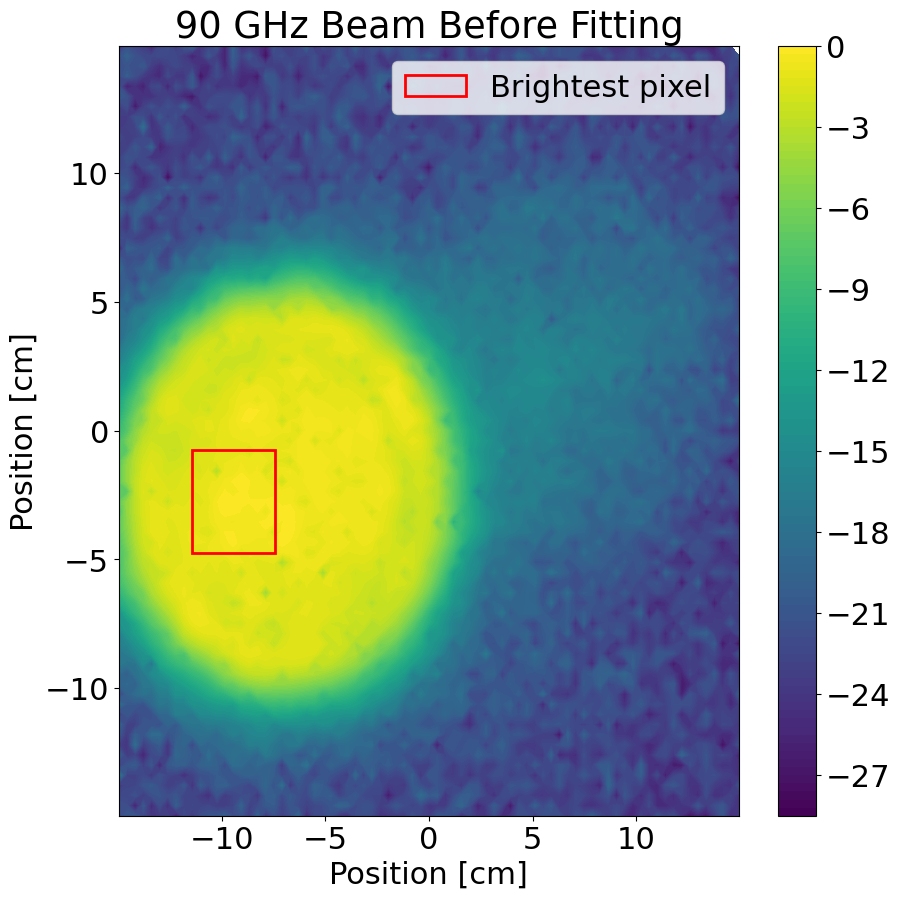}
        \includegraphics[scale=0.35]{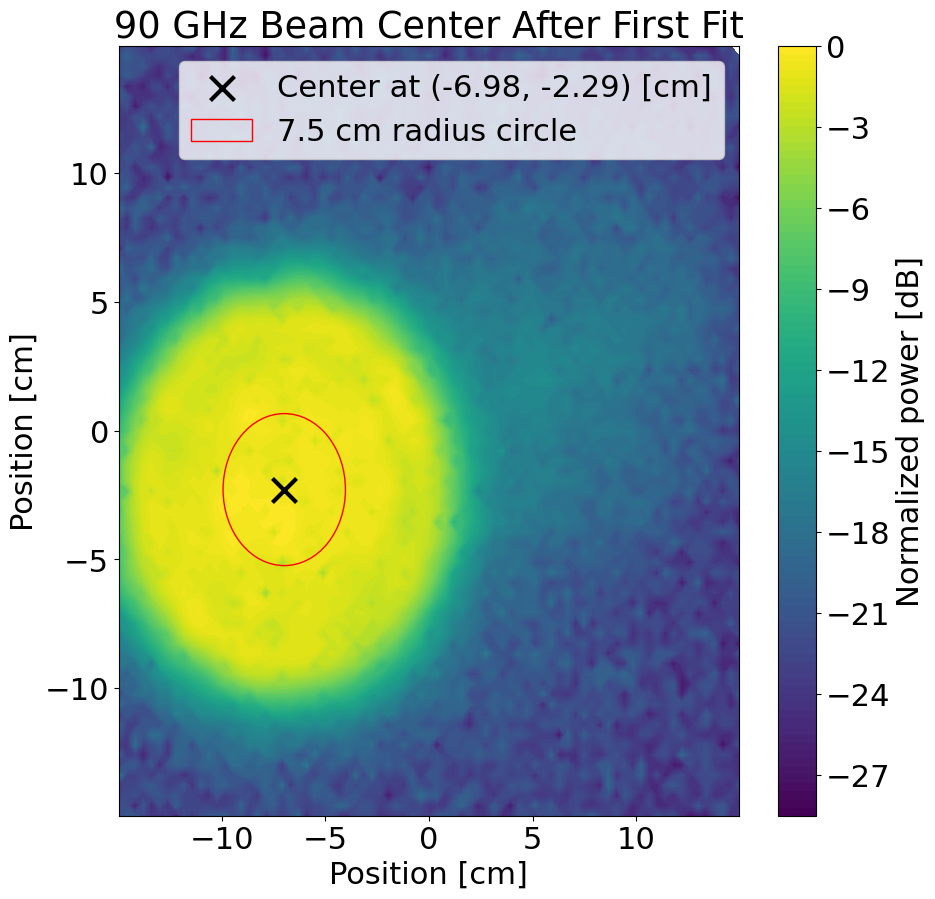}
        \caption{Left: Observed beam map before applying the first ellipse fit. The red square encompasses the brightest pixel. Right: Observed beam map after first ellipse fit. Red circle with a radius of 7.5cm depicts the area used to calculate the average signal for the final ellipse fit.
        }
        \label{fig:cleaned_beam}
    \end{figure}

\begin{figure}[ht!]
        \centering
        \includegraphics[scale=0.35]{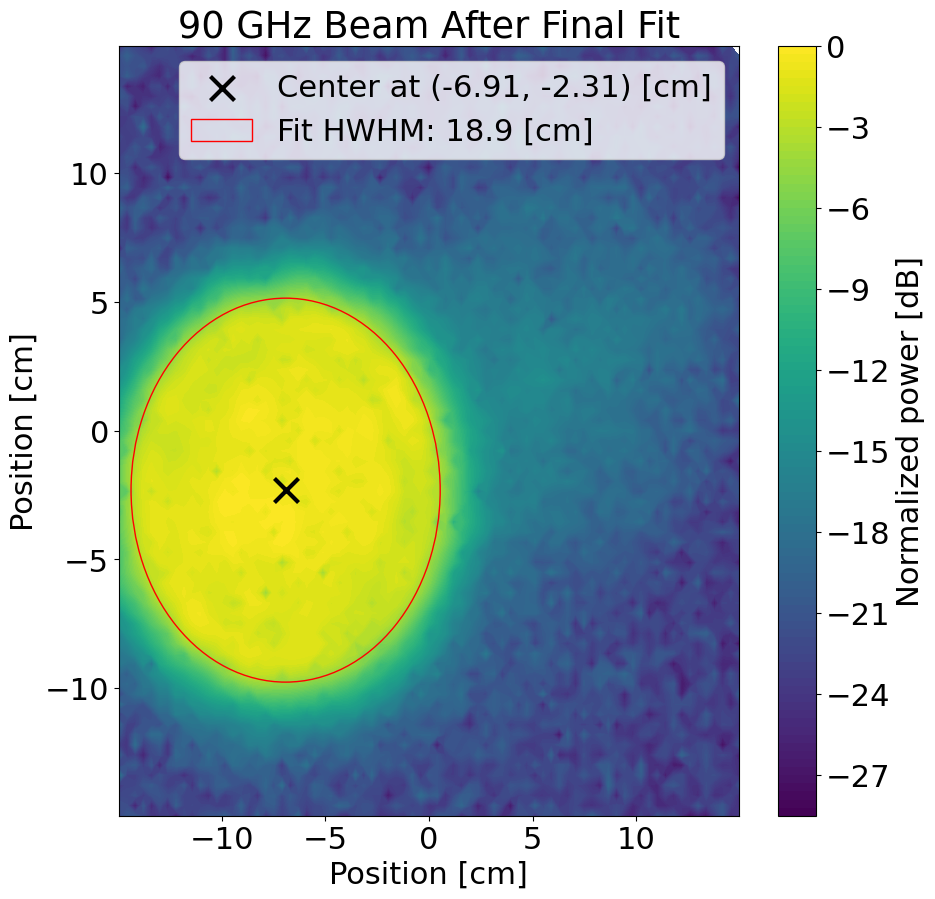}
        \caption{Observed beam map after applying the second ellipse fit. The red circle is the results of the fitted ellipse with the major axis as the radius.}
        \label{fig:final_ellipse_fit}
    \end{figure}

    The first ellipse is fit over all pixels that observe a signal power $\ge -3 dB$ from the maximum signal in the beam map. The left panel of figure ~\ref{fig:cleaned_beam} approximates the location of the maximum signal pixel (red box). We then calculate the average power within a 7.5cm radius circle around the fitted ellipse centroid (figure ~\ref{fig:cleaned_beam}, right), and re-fit an ellipse to all pixels that observe at least half of the average power around the center of the beam (figure \ref{fig:final_ellipse_fit}). The averaging smoothes the observed beam and provides a more accurate location of the center of the beam. Thus, we use the center calculated from the second ellipse fit as the center of the observed beam in the 90GHz passband detectors.   

\FloatBarrier

\FloatBarrier

\acknowledgments 
This work was supported in part by a grant from the Simons Foundation (Award \#457687, B.K.). JEG acknowledges support from the European Union (ERC, CMBeam, 101040169).
 

\bibliography{report} 
\bibliographystyle{spiebib} 

\end{document}